\documentstyle[preprint,aps]{revtex}

\tightenlines
\begin{document}
\draft
\input{epsf}
\preprint{HEP/123-qed}
\title{A possible way to look at the last and future events for two-level system}
\author{Evgueni V. Kovarski}
\address {ekovars@netscape.net}
\date{07.06.2001}
\maketitle
\begin{abstract}
In view of a three-dimensional picture (3D) of probability to find a particle
at a plane of the frequency and the time (PTF) becomes that process of absorption
and process of radiation for two - level system have different direction on the time.
Both processes are in the past or in the future depending on named
transition due to reversible model of the two-level atom.
The opportunity to know the last history of the resonant
event for the absorption process is for quantum interference interaction or for
the resonant radiation. On the contrary, to predict the resonant event in the future is possible
only by use near resonant atom-field absorption or by use radiation at a fixed time.
The problem of life time for a particle
is entered through time of spontaneous radiation connected with trajectory of the quantum transition.
It is offered to connect a trajectory of a particle during quantum
transitions with distribution of probability to find a particle.
   The conception of the spectral history for  events due to
probability distribution is introduced for next discussion.

\end{abstract}


\narrowtext

 The probability  to find a particle on the upper energy level within  two -level system
is the function of two variables such as the time $\tau$ and the frequency $\omega$.
We consider $ \it{P}(\tau,\omega)$ as the function of  probability in system of
axis coordinates  $\tau$ and $\omega$, where we shall to discuss 3D pictures of
probability - time - frequency ($\it{PTF}$).
Such images underlie present work. On them the connection between events is well visible.

 The function $ \it{P}(\tau,\omega)$ of probability is a distribution,
which has direct connection with observable structures of spectral lines and
that reminds similar distribution of the $\it{EM}$ signal theory ~\cite{Cohen}.
Between these performances of processes exists the problem of recognition for events which occurs
in an atom, but are registered in the laboratory ~\cite{E1}.
Real model can be well imagine if we shall stop more in some physical parameters.
For the near resonant absorption problem is usually difficult to imagine why there is an absorption, if there is no exact resonance.
It is necessary well to imagine where are frequencies borders of such interaction.
The account of spectral width concerns also to this. This is imortant only for long pulses
or continuous wave $\it{(cw)}$ - lasers.
Besides to believe in a nature of spontaneous radiation in system not having interactions
with anything, except with a field, is difficult. This concerns to no radiating transitions too.

   The Rabi frequency $\Omega_{\nu}$ measured in ($\it{MHz}$) is:

\begin{equation}
\Omega_{\nu} = \frac{\it{d}_{1,2}\cdot\it{E}_{0}}{\hbar}  \\
\end{equation}
where $\it{d}_{1,2}$ - is the matrix dipole momentum and $\it{E}$ - is the amplitude of EM wave.
There is the frequency $\Omega$, that make a round oscillations between two levels.
Each round consist on well known processes of the absorption and the emission due to cw - laser excitation with the field
\begin{equation}
\it{E}=\it{E_{0}}\cdot\it{cos}(\omega\cdot\tau)
\end{equation}
Therefore, we can consider only the absorption process or the radiation process with the
frequency measured in ($\it{rad./s}$): $\Omega = 2\pi\Omega_{\nu}/2=\pi\Omega_{\nu}$.
We shall notice, that considering Rabi frequency as varied parameter, we should operate
intensity $\it{I}$ of the cw-laser in space of laboratory by use the known relation:

\begin{equation}
\Omega_{\nu} = \it{const}\cdot\sqrt{I} \\
\end{equation}
 The frequency tuning conditions for the quantum interference are:

\begin{equation}
 E(\tau)= E_{0} \left[\cos (\omega_{1}\tau) + \cos(\omega_{2}\tau)\right] \\
\end{equation}
\begin{equation}
\Delta\omega_2 = \omega_2-\omega_0\\
\end{equation}
where we set:

\begin{equation}
\Delta\omega_1 = \omega_0-\omega_1 \\
\end{equation}
For the quantum interference perturbation of the upper energy level $ \it{W}_{2}$ [Fig.~1], there
are two symmetrical frequencies, so we can use the relation for exact solution:

\begin{equation}
\Delta\omega_{2}=\Delta\omega_{1} =\Delta\omega \\
\end{equation}
Let's once again address to the well known  process of an atom-field interaction,
where for events with a particle exist a well known
formula for the probability, containing the Rabi frequency $\Omega$:

\begin{equation}
P_{1}=\left[\frac{ 4\Omega^2}{ 4\Omega^2+ \Delta\omega^2}\right]
\sin^2 \left(\frac{\tau}{2}\sqrt{4\Omega^2 +\Delta\omega^2}\right) \\
\end{equation}
The transition probability for the quantum interference effect is ~\cite{E1}:

\begin{equation}
\ P_{2}=\left[\sin\left(\frac{\Omega}{\Delta\omega}\cdot\sin(\Delta\omega\tau)\right)\right]^{2} \\
\end{equation}

The experiment condition in the visible range of a spectrum for
the Lithium atom was chosen.
The correspondence between
Rabi frequencies and fine structures at the same energy level
are shown in the [Fig.~1].
When the amplitude of the laser wave $\it{E}$ is about 8,7 $\it{Volt/cm}$
the intensity of the laser is about $\it{I}$ = 100  $\it{mW/cm}^{2}$
and $\Omega\simeq$ 1000 $\it{MHz.rad.}$ , that corresponds to the Rabi frequency
$\Omega_{\nu}\simeq$ 318 $\it{MHz}$ and the intensity
 375 $\it{mW/cm}^{2}$. Because the given value for the Rabi frequency is about 1
$\it{GHz.rad}$ , we can to compare it for Lithium, where the
distance between $\it{2 P}_{1/2}$ and $\it{2P}_{3/2}$ for two isotopes is above
($2\pi)~10\it{GHz}$ and the frequency $\omega_0 \cong 2,8\cdot 10^ {6}\it{GHz.rad.}$ .
Other real parameter is the frequency $\omega_{0}$ of the laser with which is possible very precisely
to operate. For real lasers such as the dye-laser (Coherent, Model 899 -21 )
with temperature stabilized reference cell,
the frequency drift is only 50 $\it{MHz/ hour}$  with 500  $\it{kHz}$ line width.
For the quartz rod resonator structure of the broad band dye - laser, $\it{ Spectra Physics, 375 D}$ with
special procedure ~\cite{E8} for a single mode operation a resulting temperature
sensitivity was about 90 $\it{MHz}/ ^{0}\it{C}$. With  respect to the refraction
coefficient of air a temperature sensitivity is about 410 $\it{MHz}/^{o}\it{C}$.
For semiconductors lasers such as the 6202 model of $\it{New Focus}$ diode
laser or for the $\it{EOSI}$ diode laser, the line width is about 100 $\it{KHz}$ and
the average power 6 $\it{mW}$, the stability parameters at the needed wave length are better.
Therefore it is real to scan an atomic transition with accuracy about $\Delta\omega = 2 0 MHz$

 Imagining real conditions for the two -level model of Lithium atom at $(\it{2S-2P})$, look at a
3D picture of probability $\it{P}_{1}$ [ Fig.~2 ].
The probability $\it{P}_{1}$ for the quantum transition is one function as known $\it{sinc}^{2}$ and,
 as well known from the theory, it is depend on the time too, as the delta function.
When we use this probability at the small range of the time
(usually named as the probability of transition) we lose the information about the time picture
for the quantum process.  Therefore in this work we will use only $\it{P}_{1}$.
It is visible that the limit for $\it{P}_{1}= 1$ exist, because it is possible to present
that there is a spatial inclination of function of distribution.
It is a new key, which can change our performance about properties and trajectory of a particle ~\cite{E1} .
The nature of a spatial inclination $\it{P}_{1}$ far from a resonance  can be
connected with the movement of a particle along the time axis at different frequencies
at each level. A new concept for a history of events within the two-level system from here follows.
The inclination can occur for two reasons.
Firstly, the particle at one level goes along an axis of time with the greater speed,
than on the friend. It can result in electrical distribution of charges in time.
Secondly, that the particle goes with identical speed, but its way upward or downwards
differs from the standard model.
We note that for lowest energy level the pictures differences by the phase.
Therefore a direction of time for return transitions (the radiation) will differ is familiar.

  Such transformation can be applied to both the understanding of a trajectory of a particle
and the life time ~\cite{E1}. Let us to receive the possible relation between the spontaneous emission time
$\it{t}_{s}$ and the life time  $\it{t}_{L}$ of the particle within a two -level atom
having in view of a question about to define them in the time axis.
The time of spontaneous emission $\it{t_{s}}$ is well known:

\begin{equation}
t_{S} = \frac { 3\pi\hbar\epsilon_{0} c_{0}^3 }{\omega_{0}^3d_{21}^2}    \\
\end{equation}
where for an atom of Lithium some important parameters are:
$\it{t}_{S}$ = 27 ,1 $\it{ns}$,
$\omega_{0}\simeq(2\pi) 4,468\cdot10 ^ {14} \it{s}^{-1}$ ,
the dipole momentum is $\it{d}_{21}$ =2,3452
$\it{a.u.}$  or $\it{1 ,988 }\cdot10^{-29}\it{C.m.}$.
The spontaneous emission time for the Lithium atom is
$\it{t}_{S}$ = 27,1 $\it{ns}$. It causes that the system decays with a
damping constant $\it{\gamma}_{S}= \it{t}^{-1}_{S} $ when the EM field is switched off.

Now we shall to explain new definition of the life time $\it{t}_{L}$.
The probability that the particle with the life time
having unknown value $(\it{t}_{L})$ will leave the upper energy level, and its
spontaneous radiation will fade with known constant $(\gamma^{-1}_{\it{S}})$
is defined by function

\begin{equation}
\it{G}_{1}=\gamma_{S}\cdot\it{exp}\left(-\gamma_{S}\cdot\it{t}_{L}\right)
\end{equation}
The probability that the upper level
will become empty with a known damping constant $(\gamma)_{L}$ during the
unknown time of spontaneous radiation $(\it{t}_{S})$ is :

\begin{equation}
\it{G}_{2}=\gamma_{L}\cdot\it{exp}\left(-\gamma_{L}\cdot\it{t}_{S}\right)
\end{equation}
Both functions $\it{G}_{1}$ and $\it{C}_{2}$  are shown in the [Fig.~3 ]
The function $\it{G}_{2}$ has the maximum, when the first function
$\it{G}_{1}$ decreased to the value $(\it{t}_{S}\it{e})^{-1}$ at $\it{t}_{L}$ =
$\it{t}_{S}$ = 27,1 $\it{ns}$ .
The life time from the first function $\it{G}_{1}$  is equal to $ 2\it{t}_{S}$
at $\it{G}_{1M}/\it{e}^{2}$.
We can propose that the extremity of the lifetime
is at the same $\it{G}_{1M}/\it{e}^{2}$.
Therefore the equation is:

\begin{equation}
\it{X} = \it{exp}(\it{X - 2})
\end{equation}
where $\it{X}= \it{t}_{S}/\it{t}_{L}$.
The function $\it{G}_{2}$  has two interesting solutions  for the
lifetime $\it{t}_{L2}= \it{t}_{S}\cdot(20/\pi^{\star})$ ,
$\it{t}_{L1}=\it{t}_{S}/(\pi^{\star})$
We will use the relation
between the life time  $ t_{L}=\gamma_{L}^{-1}$  and
spontaneous emission time $ t_{S}=\gamma_{S}^{-1}$
Accuracy is about $(\pi^{\star}=\pi \pm 0.0045)$ :

\begin{equation}
\gamma_{S}= (19 /\pi^{\star} )\gamma_{L} = (6 ,048 ) \Gamma_{L}   \\
\end{equation}
where for $\gamma_{S}= 36,9 \it{MHz}$ one can easy obtain
$\it{g}_{L}=6,1 \it{MHz}$ , that is the quite value of the FWHM for
the $\it{2S-2P}$ Lorentz spectral line.
Also there are two options for introduced lifetime damping coefficient
$\gamma_{L}$. The first damping coefficient may be  equal
$\Gamma_{Lg}=\pi^{\star}/20 \it{t}_{S}= 5,796 \it{MHz}$.
The second is $\gamma_{L}=\pi^{\star}/19 \it{t}_{S}= 6,101 \it{MHz}$, see [Fig.~3 ].
 We note that the solution
 $20/\pi^{\star}$ is close with accuracy 0,083 to the value of $2\pi$, that is
 usually used for the frequency scale in radians. However
 such coefficients are better,
 then $ 2\pi$ due to the possible confusion for Lithium
 $\gamma_{S} = 2 \pi\cdot\gamma_{L}$.
We think that introduced relations are important
 for the Lithium atom model, because exact values are needed for the theory and experiments.
 At definition of complete probability for the life time
it is necessary to take into account both probabilities [ Fig.~4 ]:
\begin{equation}
\it{G}=\it{G}_{1}\cdot\it{G}_{2}=\frac{1}{\it{t}\cdot\it{t}_{S}}\cdot
\it{exp}\left[-\frac{\it{t}^{2}_{L}+\it{t}^{2}_{S}}{\it{t}_{L}\cdot\it{t}_{S}}\right]
\end{equation}
We note that the lifetime $\it{t}_{L}$  going from function $\it{G}_{2}$ is oversized the
lifetime $\it{t}_{L} = 2 \it{t}_{s}$  from the first distribution $\it{G}_{1}$
due to possible contribution of no radiative decay.
We assume, that it is connected to features of a trajectory of a particle
and in this connection with change of speed (acceleration or braking),
there is a problem about structure of atom ~\cite{E1}.

  Both distributions $\it{P}_{1}$ for a single
frequency tuning probability and $\it{P}_{2}$ for a probability of a quantum interference can be written
without mixture effects due to influence of the power broadening by the laser intensity and, therefore,
without value of the Rabi frequency $\Omega_{\nu}$  in equations for the
probability $\it{P}(\Delta\omega,\it{t})$. The introduced axis transformation for 3D pictures are:

\begin{equation}
\tau\cdot\Omega = X  \\
\end{equation}
\begin{equation}
\Delta\omega = Y \cdot \Omega  \\
\end{equation}
where both the $\it{X}$ and the $\it{Y}$ are linearly variable numbers.
Once again we shall pay attention that we do not know as time actually varies.
On the contrary the way of changes for a frequency of the laser is well known in each case,
therefore with approximation it is possible to consider this change as a linear.
At the frequency domain, one can use another time dependence not only linear was used.
For example, the presentation of the $\it{X}$ and $\it{Y}$ values are possible with the
projections from the complex dipole, when  the both axis have the
correspondence to the projections of the plane complex vector, e.g.
$\it{(sin X)}$ and $\it{(cos Y)}$, where for the upper energy level at the
initial time $\it{(t =0 )}$ the probability is zero too. It is easy to observe that there are closely
 pictures for the quantum interference excitation and  for the
 single frequency  excitation. These pictures was named atomic PTF lattices,
 due to the analogy with standing wave picture. It is interesting, but their use is improbable.

    We shall show, that in space of atom it is possible to predict events about
the future events only if to use interaction with EM field having one frequency
and it is possible to know about last events if to use quantum interference effect.
    We can see the equation of probability-time-frequency at the 3D picture
(PTF)[ Fig.~5]. We can precisely tell, that in the same time the particle
can pass from below upwards, but also it can move on one of probable ways to the
future places creating the future events. The movement upwards will be continuous
because the distribution of probability is based on trigonometrical function.
The movement on other directions will be quantum.
For a fixed time we can predict a future events. In this case it is important to
determine frequency of the laser to know in what time a particle will rise upwards.
On the other hand measuring spontaneous radiation at the given moment of time
we can count that it is that range of time, which corresponds to those PTF trajectories
whence in the past the particle has come.

    For two symmetrical frequencies  a direction of upwards events is opposite
[ Fig.~6 ]. For the quantum interference effect one can observe
other Rabi oscillations along the time axis $\it{(X)}$ with other
periods, so it is possible to speak about the future near resonant events.
Here is the connection with the theory of dressed atom.
The particle is with probability to equal unit
simultaneously from both parties from an exact resonance position at once in several
places on frequency [ Fig.~7].
There are some open problems such as the duality, fractional mass and charge.
However, if to accept, the particle comes in these places on different
trajectories, therefore such problems is not consider.

    We note that the Rabi oscillations have the same period for both
functions when the frequency tuning is zero and the time of interaction is
not too large. The intensity of the laser can transform axis of PTF.

   There is the conception of the spectral history as the possible way to
look at the last and future events for two-level system
in view of a three-dimensional picture (3D) of probability to find a particle
at a plane of the frequency and the time (PTF). It becomes that process of absorption
and process of radiation for two - level system have different direction on the time.
Both processes are in the past or in the future depending on named
transition due to reversible model of the two-level atom.
The opportunity to know the last history of the resonant
event for the absorption process is for quantum interference interaction or for
the resonant radiation. On the contrary, to predict the resonant event in the future is possible
only by use near resonant atom-field absorption or by use radiation at a fixed time.
The problem of life time for a particle is entered through time of spontaneous
radiation connected with trajectory of the quantum transition.
It is offered to connect a trajectory of a particle during quantum
transitions with distribution of probability to find a particle.
   The conception of the spectral history for  events due to
probability distribution is introduced for next discussion.

\section{Acknowledgments}
    Words of gratitude I devote to my friends and my family.

\section{Figure Caption}

Fig.1. Lithium atom $\it{2S-2P}$ transition as one example.

Fig.2. New 3D picture of the two-level atom.

Fig.3  The relation between the life time and the spontaneous emission time.

Fig.4  Definition of complete probability for the life time.

Fig.5. 3D picture of Probability - Time - Frequency for near resonant interaction.

Fig.6  3D picture of Probability - Time - Frequency for the quantum interference.

Fig.7  The spectral line for the quantum interference at a fixed time.

\end {document}